\documentclass[sigconf, nonacm, natbib=false]{acmart}
\usepackage{subcaption} 
\usepackage{graphicx}   
\usepackage{booktabs}   
\usepackage{tabularx} 
\usepackage{multirow}
\usepackage{makecell}
\usepackage{enumitem}

\AtBeginDocument{%
  }

\setcopyright{acmlicensed}
\copyrightyear{2026}
\acmYear{2026}
\acmDOI{XXXXXXX.XXXXXXX}
\acmConference[Conference acronym 'XX]{Make sure to enter the correct
  conference title from your rights confirmation email}{June 03--05,
  2026}{Woodstock, NY}
\acmISBN{978-1-4503-XXXX-X/2026/06}

\RequirePackage[
  datamodel=acmdatamodel,
  style=acmnumeric,
  ]{biblatex}

\begin{document}

\title[]{Not All Students Engage Alike: Multi-Institution Patterns in GenAI Tutor Use}


\author{Youjie Chen}
\affiliation{%
  \institution{The University of Hong Kong}
  \city{Hong Kong}
  \country{China}}
\email{yjmchen@hku.hk}

\author{Xixi Shi}
\affiliation{%
  \institution{Tsinghua University}
  \city{Beijing}
  \country{China}}
\email{sxx24@mails.tsinghua.edu.cn}

\author{Xinyu Liu}
\affiliation{%
  \institution{Tsinghua University}
  \city{Beijing}
  \country{China}}
\email{xyliu23@mails.tsinghua.edu.cn}

\author{Shuaiguo Wang}
\affiliation{%
 \institution{Tsinghua University}
 \city{Beijing}
 \country{China}}
\email{sw@tsinghua.edu.cn}

\author{Tracy Xiao Liu}
\affiliation{%
  \institution{Tsinghua University}
  \city{Beijing}
  \country{China}}
\email{liuxiao@sem.tsinghua.edu.cn}

\author{Dragan Ga\v{s}evi\'{c}}
\affiliation{%
  \institution{Monash University}
  \city{Clayton}
  \state{Victoria}
  \country{Australia}}
\email{dragan.gasevic@monash.edu}

\renewcommand{\shortauthors}{ }   

\begin{abstract}
The emergence of generative artificial intelligence (GenAI) has created unprecedented opportunities to provide individualized learning support in classrooms as automated tutoring systems at scale. However, concerns have been raised that students may engage with these tools in ways that do not support learning. Moreover, student engagement with GenAI Tutors may vary across instructional contexts, potentially leading to unequal learning experiences. In this study, we utilize de-identified student interaction logs from an existing GenAI Tutor and the learning management system in which it is embedded. We systematically examined student engagement (N = 11{,}406) with the tool across 200 classes in ten post-secondary institutions through a two-stage pipeline: First, we identified four distinct engagement types at the conversation session level. In particular, 10.4\% of them were "shallow engagement" where copy-pasting behavior was prevalent. Then, at the student level, we show that students transitioned across engagement types over time. However, students who exhibited shallow engagement with the tool were more likely to remain in this mode, whereas those who engaged deeply with the tool transitioned more flexibly across engagement types. Finally, at both the session and student levels, we show substantial heterogeneity in student engagement across institution selectivity and course disciplines. In particular, students from highly selective institutions were more likely to exhibit deep engagement. Together, our study advances the understanding of how GenAI Tutors are used in authentic educational settings and provides a framework for analyzing student engagement with GenAI Tutors, with implications for responsible implementation at scale.
\end{abstract}




\maketitle

\section{Introduction}

The rapid advancement of Generative Artificial Intelligence (GenAI) has fueled post-secondary institutions worldwide to integrate GenAI-powered tools into classroom learning \cite{Davis_2024, ucl_2023, Schwartzkoff_2025}. One prominent development is GenAI Tutors, which are conversational chatbots trained for educational purposes and embedded in learning platforms to support students’ course-related questions. These tools promise timely, individualized, and scalable learning support that instructors face challenges to provide due to limited time and attention \cite{horton2015identifying, kasneci2023chatgpt}. However, the availability of GenAI Tutors alone does not guarantee educational benefits. Even when designed for educational use, students may still engage with GenAI Tutors in unintended ways, such as finding cognitive shortcuts to bypass learning \cite{bastani2025generative, fan2025beware, stadler2024cognitive}. To this end, the educational promise of GenAI Tutors is critically dependent on how students engage with them in real-world classrooms.

As students increasingly rely on GenAI tools for homework and coursework, performance measures in assignment scores become less reliable to reflect student learning \cite{yan2025distinguishing}. Examining students' authentic engagement with GenAI tools can offer new insights into learning that performance scores alone cannot capture. First, engagement indicators derived from student-GenAI interactions are process-based, which provides a more fine-grained and dynamic view of the learning process as it unfolds \cite{lamsa2025measuring, moubayed2020student, swiecki2022assessment}. Second, both behavioral and cognitive engagement patterns can indicate the degree to which students actively participate in learning when interacting with the tool or simply offloading their work \cite{chi2014icap, wu2024using}. Third, temporal engagement patterns can shed light on students’ motivation to use the tool, such as for classroom learning or homework support after school \cite{dang2020ebb, wu2024using}. Together, the behavioral, cognitive, and temporal dimensions of engagement may indicate how students obtain learning support with GenAI Tutors. Additionally, these engagement patterns may vary between instructional contexts, such as institutional selectivity and course discipline, highlighting potential disparities in how students benefit from GenAI tools \cite{mercader2020university, rafalow2020digital}. For example, students in highly selective institutions may possess more learning resources and greater self-regulatory capacity to use technology for enhancement \cite{hansen2015democratizing,porter2006institutional}. Students in Science, Technology, Engineering, and Mathematics (STEM) fields may face incentives to use the tool that are different from non-STEM fields due to the task structure \cite{neumann2002teaching}.  Recent evidence suggests early signs that GenAI-enabled learning tools may exacerbate educational inequality \cite{beckman2025genai,henadirage2025barriers,yu2024whose}. However, to date, systematic examination of student engagement with GenAI tools in authentic classrooms and contextual variations has been limited. 

Existing studies of GenAI Tutors used in authentic classrooms consistently report high adoption rates (often exceeding 80\% \cite{liu2025understanding, russell2025unlocking}) with non-negligible levels of copy-pasting, an obvious short-cutting behavior for learning \cite{ghimire2024coding, liu2025understanding}. Beyond overall usage, existing studies document that students engage with GenAI Tutors in qualitatively different ways \cite{ghimire2024coding, liu2025understanding}. Importantly, these engagement patterns, instead of frequency of use, are shown to predict academic performance \cite{russell2025unlocking}. However, studies to date focus on a small number of courses within a single institution \cite{ghimire2024coding, lieb2024student, liu2025understanding, russell2025unlocking}. For example, most studies were conducted on STEM courses \cite{ghimire2024coding, lieb2024student, liu2025understanding, russell2025unlocking}, leaving open questions about how student engagement with GenAI Tutors varies across disciplines. As a result, it is important to validate these findings using large-scale datasets across institutions. To address this gap, we provide the first large-scale analysis to systematically examine student engagement with a GenAI Tutor across multiple post-secondary institutions.

Specifically, we leverage massive interaction log data from a learning management system (LMS) and its integrated GenAI Tutor. Because students interact with the GenAI Tutor through multiple conversation sessions over time, we first identified \textit{engagement types} at the session level, and then aggregated engagement types and identified \textit{engagement patterns} at the student level. Through this, we aimed to answer the following research questions (RQs):

\begin{description}[leftmargin=0em, labelindent=0em, itemsep=0.3em]
    \item[RQ1:] How much do students use the GenAI Tutor? Does it vary over time and across instructional contexts?
    \item[RQ2:] What distinct engagement types emerge at the conversation session-level? Do these engagement types vary across instructional contexts?
    \item[RQ3:] How do students transition across these conversation session-level engagement types over time? Do these student-level engagement patterns vary across instructional contexts?
\end{description}

We make three main contributions to the research on GenAI Tutors in education. First, our results advance a comprehensive understanding of how students engage with GenAI Tutors by utilizing large-scale, multi-institutional student-AI interaction data and examining through behavioral, cognitive, and temporal dimensions of engagement. Second, we propose a learning analytics framework that identifies conversation session-level engagement types and integrates with student-level engagement patterns, which can be applied to other human-AI interaction data. Third, our analysis of contextual variation at both the session and student level reflects equity concerns about the use of GenAI in education settings. Insights from this study can inform the development of learning analytics interventions and guide GenAI regulations in institutions to support student learning with GenAI Tutors in more effective and equitable ways.

\section{Related Work}
\label{sec: Related Work}

\subsection{The Evolution of Automated Tutoring Systems}
\label{sec: Automated Tutoring Systems}

The development of automated tutoring systems is motivated by the pursuit of the “2 sigma” effect, which refers to the finding that students perform two standard deviations better when receiving one-to-one human tutoring compared to traditional classroom instruction \cite{bloom19842}. For decades, Intelligent Tutoring Systems (ITSs) have been developed to mimic one-to-one human tutoring by building four core modules to provide personalized tutoring: domain, student, pedagogical, and interface modules \cite{anderson1985intelligent, ma2014intelligent,mousavinasab2021intelligent}. With proper scaffolding designed in the system, ITSs have been shown to improve learning outcomes nearly as effectively as one-to-one human tutoring in experimental studies \cite{ma2014intelligent,vanlehn2011relative}. However, the complexity of the system architecture often makes it costly to develop and maintain \cite{anderson1995cognitive,mousavinasab2021intelligent}, and the effectiveness of the system cannot easily be generalized across subject domains \cite{feng2021systematic,slavuj2015intelligent,zhang2017evaluating}. As a result, despite notable advances in improving student performance outcomes, scaling ITSs to diverse educational contexts remains challenging (for exceptions, see \cite{pane2014effectiveness,pardos2023oatutor}).

Recent advancements in Large Language Models (LLMs) have fueled the development of LLM-enabled automated tutoring systems, specifically GenAI Tutors. Unlike ITSs, GenAI Tutors are lightweight conversational agents that typically lack structured architectural models of domain, student, and pedagogy \cite{pal2024autotutor}. Instead, they rely on API-based LLMs to interact with students in natural languages, with additional training and prompt engineering tailored for educational use \cite{johnson2024harness,liu2025understanding,schmucker2024ruffle}. Due to the availability of LLM APIs, GenAI Tutors can provide learning support at lower costs than those associated with ITSs development \cite{li2025bringing,yan2024promises}. Their natural language delivers a more positive user experience \cite{lieb2024student} and requires less cost for student or teacher training on how to use the tool. Additionally, the same GenAI Tutor can be used across different subject domains, as the underlying LLMs are trained on broad information instead of data in a focused discipline \cite{li2025bringing,pal2024autotutor}. As a result, this creates advantages for it to scale compared to ITSs.

However, GenAI Tutors face challenges in providing effective learning support. Prior experiments find that the effects of GenAI Tutors on learning outcomes are mixed \cite{bastani2025generative,brender2024s,lehmann2025aimeetsclassroomlarge}. For example, Lehmann et al. \cite{lehmann2025aimeetsclassroomlarge} show that GenAI Tutors improve learning when students use them to seek explanations, but they can harm learning when students over-rely on them to generate solutions. Borchers and Shou \cite{borchers2025can} demonstrate that current LLMs alone have limited instructional adaptivity and pedagogical soundness, and are unlikely to produce similar learning benefits compared to ITSs. Bastani et al. \cite{bastani2025generative} show that GenAI Tutors need to be developed with guardrails to protect students from misuse. Building on prior work, this study focuses on student engagement to reveal the potential learning support and challenges that GenAI Tutors may provide.

\subsection{Student Engagement with GenAI Tutors}
\label{sec: Student Engagement}

Recent research on student engagement with GenAI tutors reports high adoption rates but variation in usage intensity across studies. For example, Russell et al. \cite{russell2025unlocking} deployed a GenAI Tutor in an introductory chemistry course and found that 93\% of 660 participants interacted with the GenAI Tutor at least once during the semester. However, among adopters, usage intensity was relatively low, with students sending 4.6 prompts on average. In contrast, students in a business course averaged 184 prompts per student among 44 students \cite{salminen2024using}. Additionally, usage intensity can be highly skewed among students within the same course. For example, McNichols et al. \cite{mcnichols2025study} reported that low-interaction users, measured by their total number of prompts in one semester, averaged only 3.2 prompts, while high-interaction users averaged 284 prompts.

Beyond simple adoption and usage intensity measures, several studies further identified student engagement patterns \cite{brender2024s,lai2025leveraging,liu2025understanding,russell2025unlocking}. For instance, Brender et al. \cite{brender2024s} categorized 397 student prompts from a robotic course into solution development, task or concept understanding, and code debugging, and then clustered 38 students into three types: Conceptual Explorers, Practical Developers, and Debuggers. Their results show that Debuggers’ use of ChatGPT improved performance but not learning, whereas Conceptual Explorers’ use improved learning but not performance.
In another study, Russell et al. \cite{russell2025unlocking} identified seven homework problem-level interaction patterns and then grouped 602 students into five distinct clusters based on the problem-level cluster traces. They found that high-performing students used the GenAI Tutor selectively, whereas low-performing students tended to seek GenAI Tutor assistance before attempting to solve the problems themselves. In an introductory statistics course, Lai et al. \cite{lai2025leveraging} showed that high-performing students engaged more frequently in reflective and evaluative actions, while low-performing students relied more on reactive, answer-seeking strategies. In summary, existing research provides early evidence that students engage with GenAI Tutors in distinct ways that may suggest different purposes of use and different levels of cognitive engagement in discipline-specific contexts.

\subsection{Learning Analytics to Reveal GenAI Interaction Patterns}
\label{sec: Learning Analytics}

As GenAI-enabled tools become increasingly adopted in education, various learning analytics approaches have emerged to investigate student–GenAI interaction data \cite{khosravi2025generative,yan2024generative}. To identify different types of student engagement, existing studies have mostly used qualitative coding to categorize student prompts or GenAI responses following well-established or self-developed frameworks \cite{brender2024s,liu2025understanding,mcnichols2024can,orozco2025emergent}. Other work applies data-driven approaches, such as clustering techniques \cite{russell2025unlocking} or epistemic network analysis \cite{lai2025leveraging,yang2025ink}, to examine learning interactions, which can be better adopted in large-scale GenAI interaction contexts compared to qualitative coding. 

Additionally, because student-GenAI interactions unfold over time, process mining is also a suitable technique in this context \cite{chen2025unpacking,fan2025beware,urban2025prompting}. In digital learning contexts prior to GenAI interactions, process mining is extensively used to reveal how student interaction patterns unfold over time and reflect the underlying learning strategies that drive learning outcomes \cite{fan2021learning,jovanovic2017learning,matcha2019analytics,saint2018detecting}. Typically, raw interaction logs are first segmented into discrete sessions and identified as learning tactics. Then process mining is applied to aggregate these session-level learning tactics into student-level learning strategies \cite{matcha2019analytics,saint2018detecting}. Building on prior research, we propose a learning analytics framework that develops a two-stage pipeline to identify \textit{engagement types} at the conversation session level through clustering analysis and then examines \textit{engagement patterns} at the student level over time through process mining.

\section{Methods}
\subsection{Study Context}
\label{sec: Study Context}

We analyzed existing, de-identified interaction logs from a commercial LMS that is widely used in several post-secondary institutions, along with its integrated GenAI Tutor. The GenAI Tutor was launched in the LMS in Fall 2024 and has been used by more than 10,000 classes across 573 institutions via its LMS during Fall 2024 and Spring 2025. This study involves secondary analysis of existing commercial data and was approved by the Institutional Review Board of Tsinghua University (ID: THU-04-2025-1105). The research team has no access to identifiable information about students, instructors, or institutions. Only anonymized demographic attributes were available at the institution-level, as shared by the company.

This GenAI Tutor is generalizable to GenAI Tutors that are typical of the standard industrial trend: It has a technical architecture that is powered by API-based LLMs with additional engineering for educational use; then, as a feature, it is displayed as a conversational chatbot integrated into an existing educational platform. Specifically, this GenAI Tutor is built on a three-tier technical architecture \cite{Pearson_2025}. The bottom layer serves as a model hub that integrates approximately 30 distinct LLMs (including GPT-4o via Microsoft Azure, DeepSeek-R1, and GLM-4-Plus). The middle layer functions as a domain knowledge engine, which retrieves content from instructor-vetted course materials (such as lecture slides, syllabi, and textbooks) to ensure more accurate and up-to-date information than general LLMs. The top layer is the application interface integrated into the LMS, providing students with immediate, context-aware support for course-related questions.

Instructors gain access to the GenAI Tutor when their institution purchased the provider's LMS. Instructors can then make the GenAI Tutor available to their students. Instructors can tailor the GenAI Tutor to their classes by selecting the underlying LLM and uploading course materials, such as syllabi and lecture slides. Instructors and students' use of the GenAI Tutor is completely voluntary. Some universities offer training workshops on the use of the LMS and its GenAI Tutor, while others rely on self-guided tutorials and handbooks. Typically, instructors use the LMS as a complementary tool to existing LMSs from their institutions. Most institutions operate on a basic LMS developed by themselves, with basic functions such as distributing materials and managing assignment grades, while the LMS in this study is used to provide richer interactive learning activities and instructor dashboards with summary statistics \cite{wang2018rain}. Consequently, instructors commonly use their institution's LMS to manage assignments and grades, and turn to the present LMS for supplementary instructional support, such as pre-class slides and quizzes. As a result, this study does not have access to students' official course grades. 


\subsection{Data}
\label{sec: Data}

In large-scale deployment, student engagement with the GenAI Tutor varies widely. To ensure sufficient interaction data for a meaningful analysis, we focused on the ten institutions with the highest median adoption rates across their classes and 20 classes within each institution with the highest adoption rates during the Spring 2025 semester (mid-February to mid-July). We used the Spring 2025 semester data to capture students' stable usage patterns after potential novelty effects following the Fall 2024 launch. 
This yielded de-identified interaction records from 11,406 students across 200 classes at 10 institutions.\footnote{The student count represents student-class enrollments, including 10,629 unique students, among whom 679 (6.4\%) were enrolled in multiple classes (2-3 courses) within our sample. All subsequent analyses were conducted at the student-class level to account for variation in student engagement across different courses.} 

The dataset contains interaction records at both the instructor and student levels. At the instructor level, it captures information about instruction activities (e.g., class start times and end times) and exercise quizzes (e.g., number of problems, release times, and deadlines). At the student level, it contains both LMS interaction logs (e.g., if an exercise quiz is completed and the score) and GenAI Tutor interaction logs (e.g., student prompts, Tutor responses, and timestamps). The data also includes administrative information at the class level (e.g., course names and course-affiliated departments) and the institution level (e.g., if the institution is categorized as a highly selective institution\footnote{Institutional classification follows the Ministry of Education's official categorization system.}). Among the ten institutions, three are classified as highly selective universities. The 200 classes had an average class size of 57.03 (SD = 39.23) students. A total of 89 (44.5\%) classes belong to STEM. The dataset does not contain any demographic information for students or instructors.

\subsection{Analytical Approach}
\label{sec: Analytical Approach}

Because each student could interact with the GenAI Tutor through multiple conversation sessions, we developed a two-stage analytical pipeline to examine engagement behaviors first at the conversation session level and then aggregated them to the student level. First, we used prediction models to identify and segment conversation sessions from raw student-GenAI interaction logs. Building on this segmentation, we examined students' adoption and use of the GenAI Tutor over time through descriptive analysis (RQ1). We then used linear regression models to assess variations across institution selectivity and course disciplines. Given that 6.4\% of students were enrolled in multiple classes, we clustered standard errors at both the student and class levels. Second, to identify session-level \textit{engagement types} (RQ2), we conducted feature engineering to characterize the conversation sessions, and then applied clustering analysis to categorize distinct engagement types. We used linear regression models to examine how these types varied across instructional contexts. Last, to analyze student-level \textit{engagement patterns} (RQ3), we applied process mining to analyze how session engagement types transitioned as students interact with the GenAI Tutor over time. We also compared transition patterns across instructional contexts by applying process mining separately to student subgroups. The full implementation is available at \url{https://github.com/VinceLiu05/edu-analysis}.

\subsubsection{Identifying and Segmenting Conversation Sessions}
\label{sec: Session Segmentation}

Here, we define a conversation session as a sequence of consecutive conversation turns between a student and the GenAI Tutor focused on a single topic within a bounded period of time. Because GenAI Tutors rely on students to initiate and terminate conversations, the start and end of these conversation sessions were not well-defined. First, we proposed the following two complementary algorithms to segment student–GenAI Tutor conversations into sessions: (1) \textit{Time-based algorithm}: The algorithm determines that a new session starts when a student’s prompt occurred more than \textit{t} minutes after the previous prompt. (2) \textit{Topic-based algorithm}: An LLM segments conversation sessions based on the semantic content of student prompts and GenAI Tutor responses.

Then, to train and validate the above algorithms, we constructed a validation dataset using student-GenAI interaction logs that occurred in specific learning activities in the LMS (e.g., a quiz problem or a slide). The interaction records on these pages served as natural approximations of topic-bounded conversation sessions and represented 8.4\% of all conversation logs, while most of the interaction occurred on the main menu page. We fine-tuned the above algorithms (e.g., adjusting the time threshold \textit{t}, the LLM prompt, and how the two algorithms were combined) on the remaining dataset and bench-marked their performance against this validation dataset using F1-scores. The algorithm first segmented sessions using a 15-minute inactive threshold and then used an LLM (Gemini-2.5-Flash) to further segment sessions based on the topic.\footnote{The prompt details for the LLM can be found in our GitHub repository: \url{https://github.com/VinceLiu05/edu-analysis}.} In the end, this approach achieved an F1-score of 0.9 and segmented 188,463 conversation turns into 113,255 conversation sessions, with a median of 5 (SD = 27.09) sessions per student.




\subsubsection{Feature Engineering}
\label{sec: Feature Engineering}

After identifying the conversation sessions, we conducted feature engineering at the session level to characterize student engagement by behavioral, cognitive, and temporal dimensions. Although engagement frameworks often distinguish behavioral, cognitive, and emotional dimensions \cite{fredricks2004school}, we did not characterize emotional engagement due to the lack of approaches to reliably infer emotions from our data logs. 
We followed engagement measurements in digital learning environments \cite{dewan2019engagement,henrie2015measuring, fredricks2004school} to capture behavioral and cognitive engagement through student-GenAI interaction logs.
We also incorporated temporal engagement, as prior research has shown that engagement patterns over time are important to understand learning behaviors \cite{jovanovic2017learning,reimann2009time,Sher2022When}. In total, we engineered 10 features that capture behavioral (N = 3), cognitive (N = 3), and temporal engagement (N = 4). Table~\ref{tab:features} presents the engagement dimensions, features, and their descriptions on how they are measured. For full details on how these features are computed, please refer to our GitHub repository: \url{https://github.com/VinceLiu05/edu-analysis}.

\begin{table*}[t]
  \centering
  \small
  \setlength{\tabcolsep}{4pt}
  \caption{Features Representing Different Dimensions of Conversation Session-Level Engagement}
  \label{tab:features}
  \begin{tabularx}{\textwidth}{l >{\raggedright\arraybackslash}p{0.29\textwidth} X}
    \toprule
    \textbf{Dimension} & \textbf{Feature} & \textbf{Description} \\
    \midrule

    \multirow[t]{3}{*}{\makecell[tl]{Behavioral\\Engagement}}
      & Number of Conversation Turns
      & The total number of conversation turns \\
    & Average Time Duration per Turn
      & The average minutes spent per conversation turn \\
    & Average Word Count per Turn
      & The average number of words per student prompt \\
    \midrule

    \multirow[t]{3}{*}{\makecell[tl]{Cognitive\\Engagement}}
      & Number of Copy-Paste Events
      & The number of copy-paste events as indicated by quiz image uploads, specific keywords (e.g., ``as follows'' and ``Question x''), language patterns that likely suggest structured exam or homework text (e.g., ``A.~\ldots~B.~\ldots'' and code blocks), and excessively long prompts \\
    & Number of Direct Answer Requests
      & The number of direct answer requests as indicated by specific keywords (e.g., ``give me the answer to\ldots'' and ``what is the solution of\ldots'') \\
    & \mbox{Number of Understanding-Oriented Queries}
      & The number of understanding-oriented queries as indicated by specific keywords (e.g, ``how to understand\ldots'' and ``why does\ldots'') \\
    \midrule

    \multirow[t]{4}{*}{\makecell[tl]{Temporal\\Engagement}}
      & Week Progress
      & The number of weeks when the conversation session occurred within an academic semester \\
    & Exam Period Indicator
      & The likelihood that the conversation session occurred during the exam weeks \\
    & Time of Day
      & The hour of the day when the conversation session occurred \\
    & In-Class Indicator
      & The likelihood that the conversation session occurred during scheduled class hours \\
    \bottomrule

  \end{tabularx}
\end{table*}

\subsubsection{Session-Level Clustering Analysis}
\label{sec: Clustering Analysis}

We conducted \textit{k}-means clustering to identify session-level engagement types based on the identified conversation sessions and their features. To prepare for clustering, we applied log-transformation to features that followed a long-tailed distribution so that extreme values would not disproportionately affect the distance calculations in clustering. Then, we standardized all features using z-score normalization to ensure that features on different scales contribute equally to the algorithm. Finally, we performed dimensionality reduction using Principal Component Analysis (PCA) to address potential multicollinearity among features. We retained six principal components, which together explained 82.3\% of the total variance. We then ran \textit{k}-means clustering using the Python package scikit-learn and applied the Elbow method to select the optimal number of clusters. In the end, we identified four clusters. 

For robustness check, we reran \textit{k}-means with 50 random initializations and found that cluster assignments remained highly stable across runs (mean Adjusted Rand Index = 0.999, SD = 0.001). In addition, we included four additional features in the clustering analysis to examine the robustness of our initial set of features. Additional features were developed with richer data among classes that used the LMS intensively: the time distance from each session to the previous class, the upcoming class, the most recent assignment release, and the nearest assignment deadline. We found that both classes with richer features and classes without them yielded the same four clusters, suggesting that the initial set of features is sufficient to identify engagement types across conversation sessions.

\subsubsection{Student-Level Process Mining}
\label{sec: Process Mining}

Building on the engagement types identified from the result of the session-level clustering, we then performed process mining to characterize student-level engagement patterns. Because students interact with the GenAI Tutor across multiple conversation sessions over time, we used a First-Order Markov Model (FOMM) to characterize how students generally transition among engagement types across sessions. We restricted the analysis to students who engaged in at least one session. We represented their interaction history as an ordered sequence of sessions with various engagement types, bounded by a start and end state (e.g., [Start, Type A, Type A, Type B, End]). In the end, we obtained a transition probability matrix that revealed how students moved across sessions with different engagement types over time. We also applied the FOMM separately among students from different instructional contexts to explore potential differences in their engagement patterns.

\section{Results}
\subsection{Descriptive Results of Student Engagement}
\label{sec: Descriptive Results}

\begin{figure}[htbp]
    \centering
    \includegraphics[width=\linewidth]{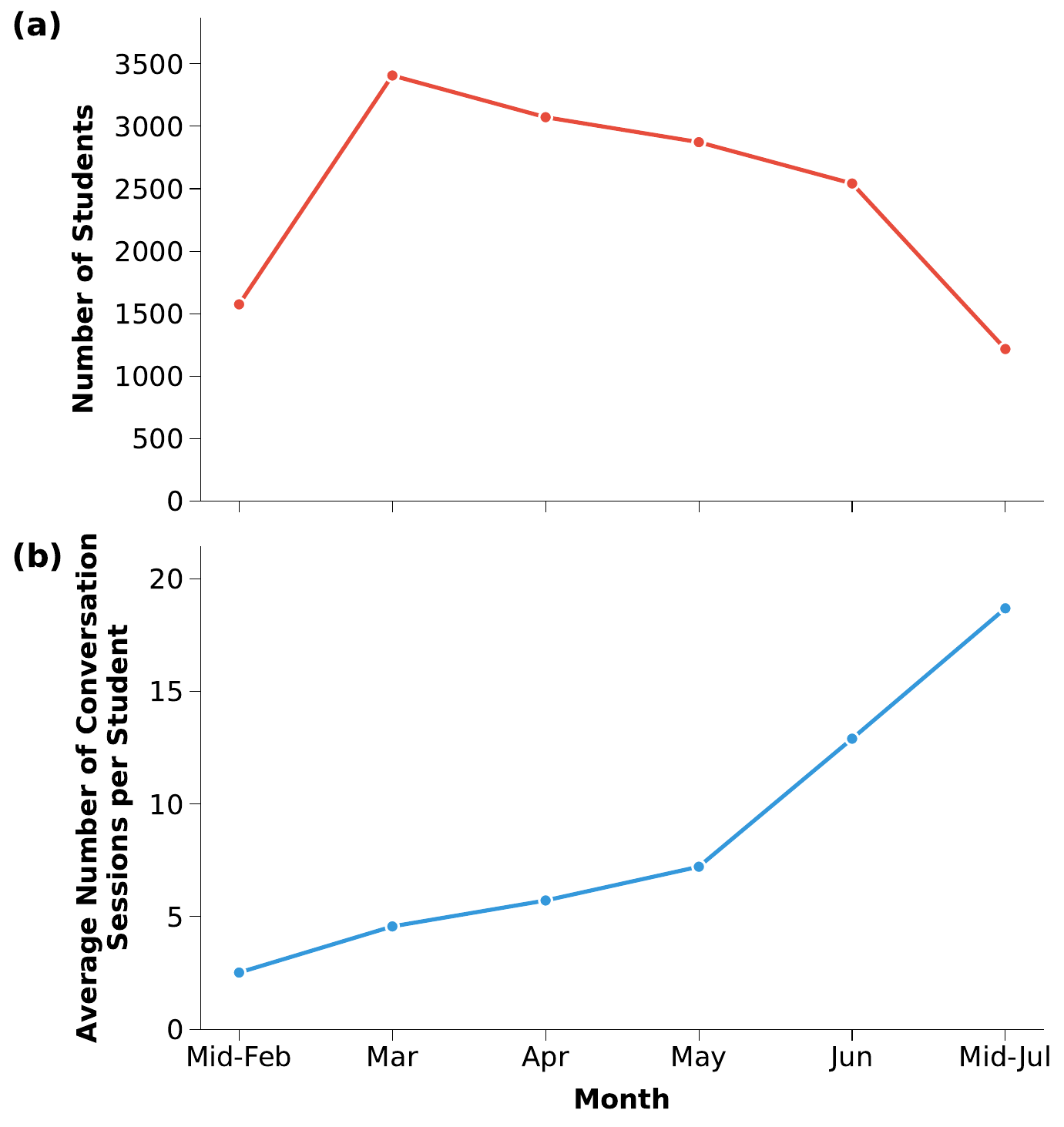}
    \caption{Monthly Trends of Student Engagement with the GenAI Tutor from Mid-February (Semester Starts) to Mid-July (Semester Ends). Panel (a) shows the number of students who used the GenAI Tutor, and Panel (b) shows the average number of conversation sessions among students who used the GenAI Tutor.}
    \label{fig:monthly_engagement}
\end{figure}

To answer the first research question, we provide students' adoption rate and usage intensity (as indicated by the number of conversation sessions) of the GenAI Tutor, as well as the temporal trend of usage intensity over the semester. Among the 11,406 students, 6,932 (60.8\%) engaged with the GenAI Tutor at least once. Among those who used the GenAI Tutor, students generated a median of 5 conversation sessions. The distribution of the number of conversation sessions is highly skewed, with a small number of students highly active in using the GenAI Tutor (25th percentile = 1; 75th percentile = 19). 
Then we examined student engagement over time. Throughout the semester, the number of students who used the GenAI Tutor peaked in the early semester and then declined (Figure~\ref{fig:monthly_engagement}.a). In contrast, the usage intensity increased steadily during the semester (Figure~\ref{fig:monthly_engagement}.b). This indicates that many students made exploratory attempts early on, while a subset of students increased their use over the semester.

Finally, we examined contextual variation in student–GenAI Tutor adoption and usage intensity. Compared to non-STEM students, those in STEM disciplines are 20 percentage points less likely to adopt the GenAI Tutor (51 \% vs. 71\%; $t(199) = -4.04$, $p < 0.01$). Among those who adopted the GenAI Tutor, students in STEM courses participated in 10.76 fewer conversation sessions on average ($t(199) = -3.48$, $p < 0.01$). In contrast, institutional selectivity was not significantly associated with adoption or usage intensity, suggesting that differences in overall engagement were closely related to disciplinary contexts but not institutional selectivity. Detailed statistical results are available at OSF: \url{https://osf.io/7jhnb/?view_only=bf8b5da0ccd4437d81323808a77b6e11}.

\subsection{Engagement Types at the Session Level}
\label{sec: Engagement Types}

\begin{figure*}[t]
    \centering
    \includegraphics[width=0.8\textwidth]{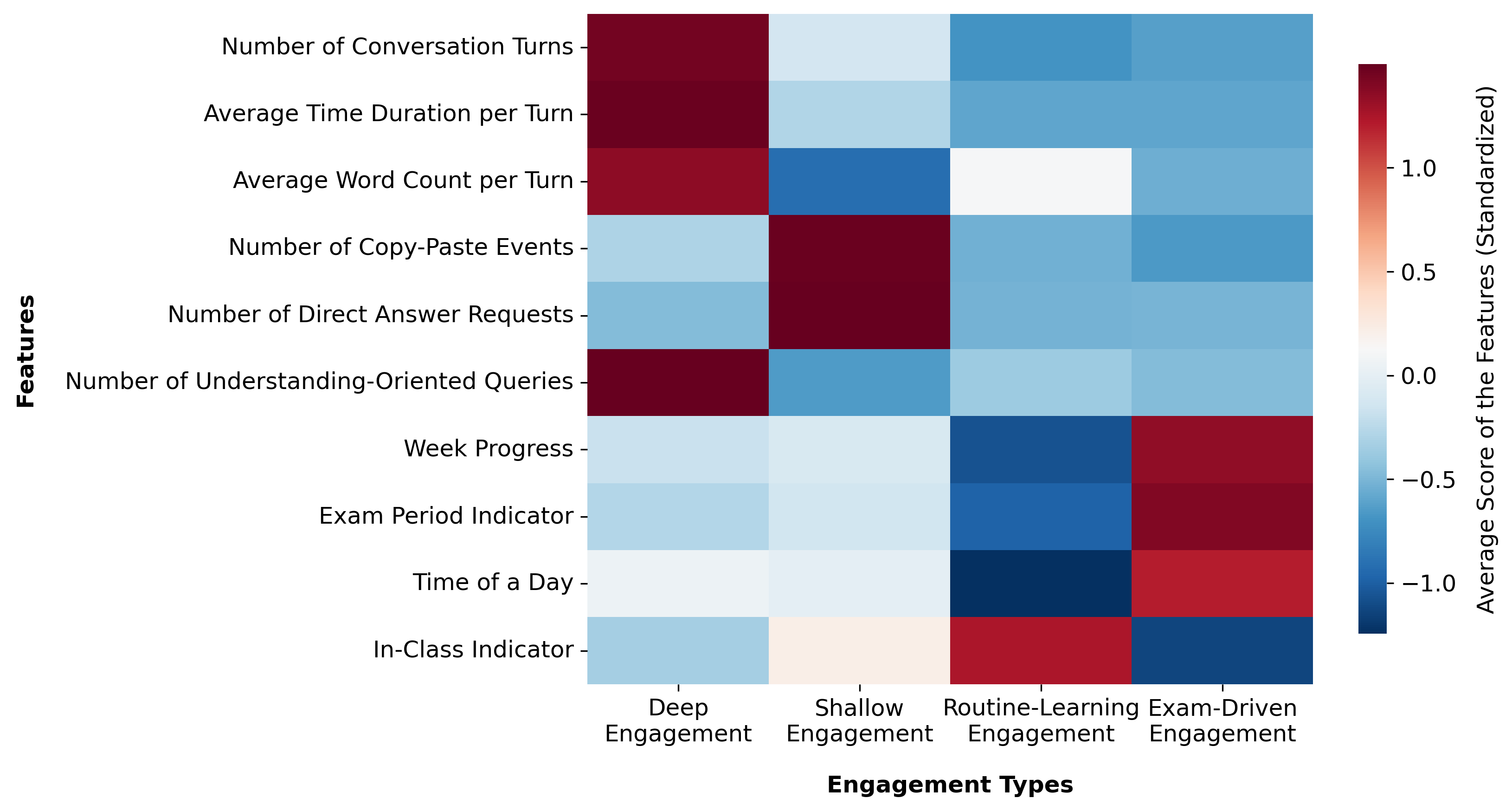}
    \caption{Heatmap of Cluster Centroids across Features. Each cell shows the standardized score (z-score) indicating how much each cluster's mean deviates from the overall mean.}
    \label{fig:cluster_heatmap}
\end{figure*}

\begin{figure}[htbp]
    \centering
    \includegraphics[width=\linewidth]{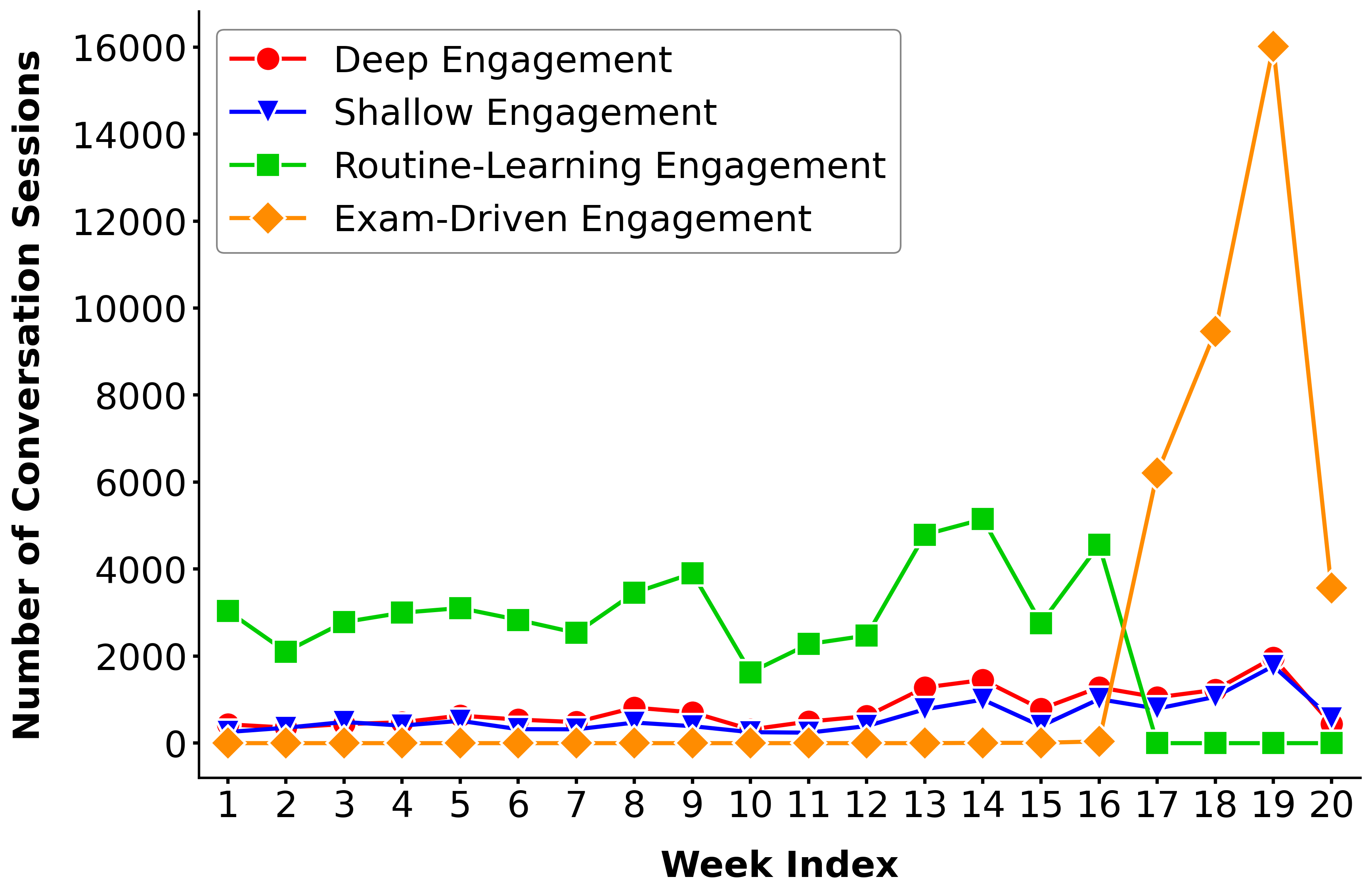}
    \caption{Distribution of Conversation Sessions across Academic Weeks by Engagement Type}
    \label{fig:term_progress}
\end{figure}

To address the second research question, we identified the following four engagement types in the 113,255 conversation sessions: Deep, Shallow, Routine-Learning, and Exam-Driven Engagement. Figure~\ref{fig:cluster_heatmap} illustrates the average standardized feature values for each type of engagement, and Figure~\ref{fig:term_progress} shows how these engagement types evolve over the semester. 

1. \textbf{Deep Engagement} (14.0\% of all sessions): These sessions exhibited high behavioral engagement, involving multiple conversation turns, tended to last longer, and included more words. They also exhibited a high prevalence of understanding-oriented queries. Temporally, they were distributed relatively evenly across an academic semester.

2. \textbf{Shallow Engagement} (10.4\% of all sessions): These sessions were characterized by few conversation turns, short interaction durations, and, in particular, few words. They exhibited shallow cognitive engagement, with a high prevalence of copy-pasting behaviors and direct answer-seeking requests. Temporally, they were distributed relatively evenly across an academic semester.

3. \textbf{Routine-Learning Engagement} (44.5\% of all sessions): These sessions occurred mainly during daytime and the first half of an academic semester, suggesting that they could be connected to routine class activities. They only had a few conversation turns, but included a reasonable amount of words in the conversations. These sessions more often involved questions aimed at understanding and less often involved direct answer-seeking requests. 

4. \textbf{Exam-Driven Engagement} (31.2\% of all sessions): These sessions were concentrated in the final weeks of an academic semester, with a pronounced increase towards the end of the semester, indicating that they could be related to exam preparation. They had a few conversation turns and were short in duration. They exhibited the least copy-pasting behaviors, indicating that students had fewer needs to obtain formalistic answers that were likely for homework assignments.

To compare behavioral, cognitive, and temporal engagement across the four engagement types, we conducted pairwise comparisons using linear regression models and yielded three main insights. First, deep engagement and shallow engagement sessions differed primarily in \textit{how} but not \textit{when} students interacted with the GenAI Tutor. Deep engagement sessions involved both more behavioral and cognitive engagement compared to shallow engagement sessions (all behavioral engagement features: $ps < 0.01$; all cognitive engagement features: $ps < 0.01$). However, temporally, the two patterns showed similar distributions throughout the academic semester, though shallow engagement sessions were more likely to occur during class (Cohen's $d$ = 0.16, $t(199) = 2.56$, $p = 0.01$) and generally less frequently than deep engagement sessions. Detailed statistical results of the pairwise comparison can be found in \url{https://osf.io/7jhnb/?view_only=bf8b5da0ccd4437d81323808a77b6e11}.

Second, routine-learning engagement and exam-driven engagement sessions were distinguished by their temporal differences. Routine-learning engagement was concentrated in the first half of the academic semester along with coursework, while exam-driven engagement was concentrated in the final weeks as exams approached. Behaviorally and cognitively, the two engagement types were broadly similar, although routine-learning engagement sessions involved more words per turn ($d$ = 0.17, $t(199) = 3.06$, $p < 0.01$) and more copy-paste events ($d$ = 0.15, $t(199) = 3.65$, $p < 0.01$) than exam-driven engagement sessions. 

Last, shallow engagement sessions can be distinguished from the two temporally-driven engagement types (i.e., routine-learning and exam-driven engagement) not just through temporal engagement, but also through behavioral and cognitive engagement. However, the effect sizes for behavioral engagement were smaller compared to cognitive engagement. For example, the effect sizes for the number of conversation turns when comparing shallow engagement were $d$ = 0.49 with routine-learning engagement ($t(199) = 11.83$, $p < 0.01$) and $d$ = 0.44 with exam-driven engagement ($t(199) = 11.45$, $p < 0.01$). In contrast, for cognitive engagement, the number of copy–paste events showed much larger effect sizes when comparing shallow engagement with routine-learning engagement ($d$ = 2.17, $t(199) = 30.57$, $p < 0.01$) and exam-driven engagement ($d$ = 2.32, $t(199) = 25.01$, $p < 0.01$).

\begin{figure*}[t]
    \centering
    \includegraphics[width=0.48\textwidth]{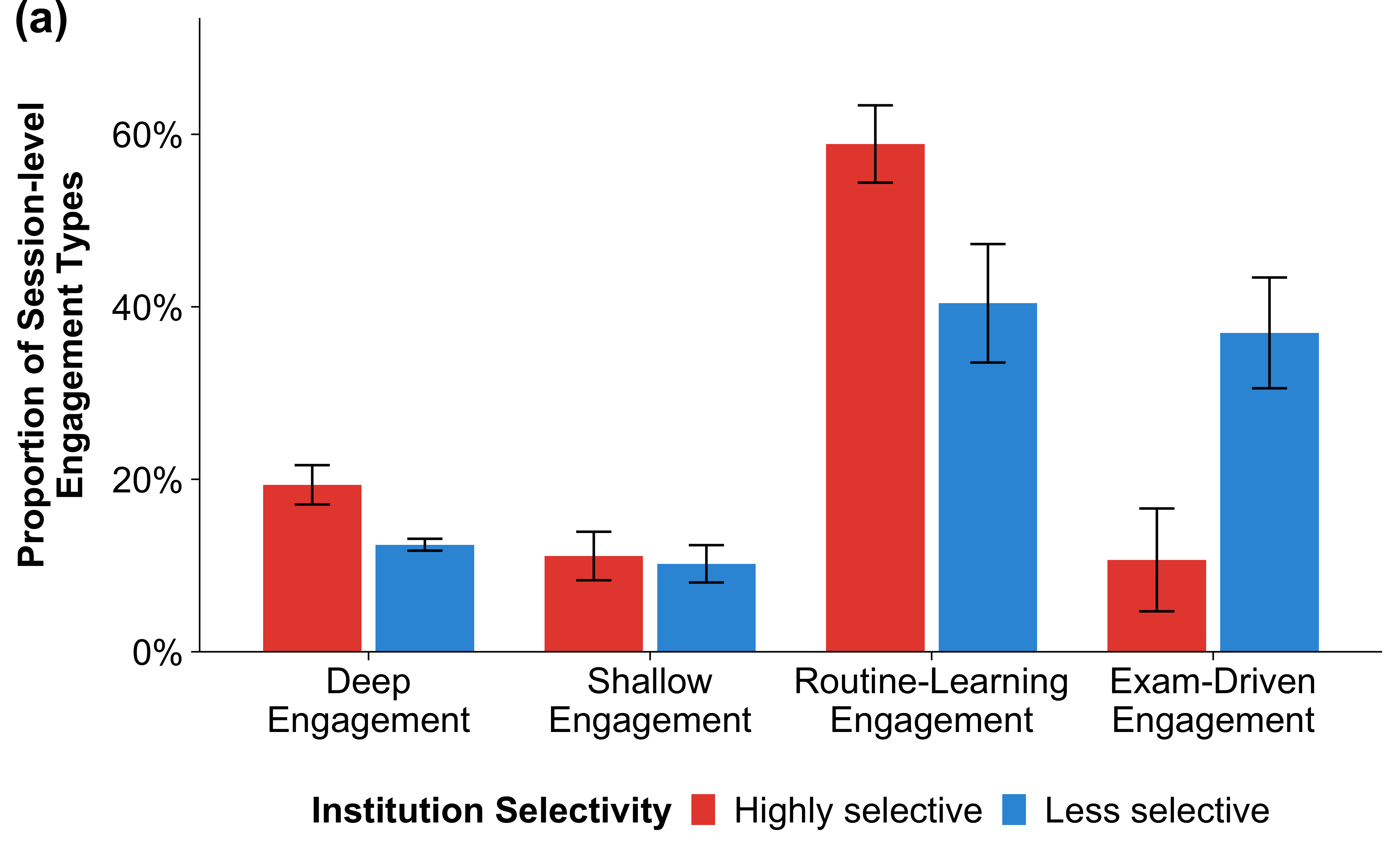}
    \hfill
    \includegraphics[width=0.48\textwidth]{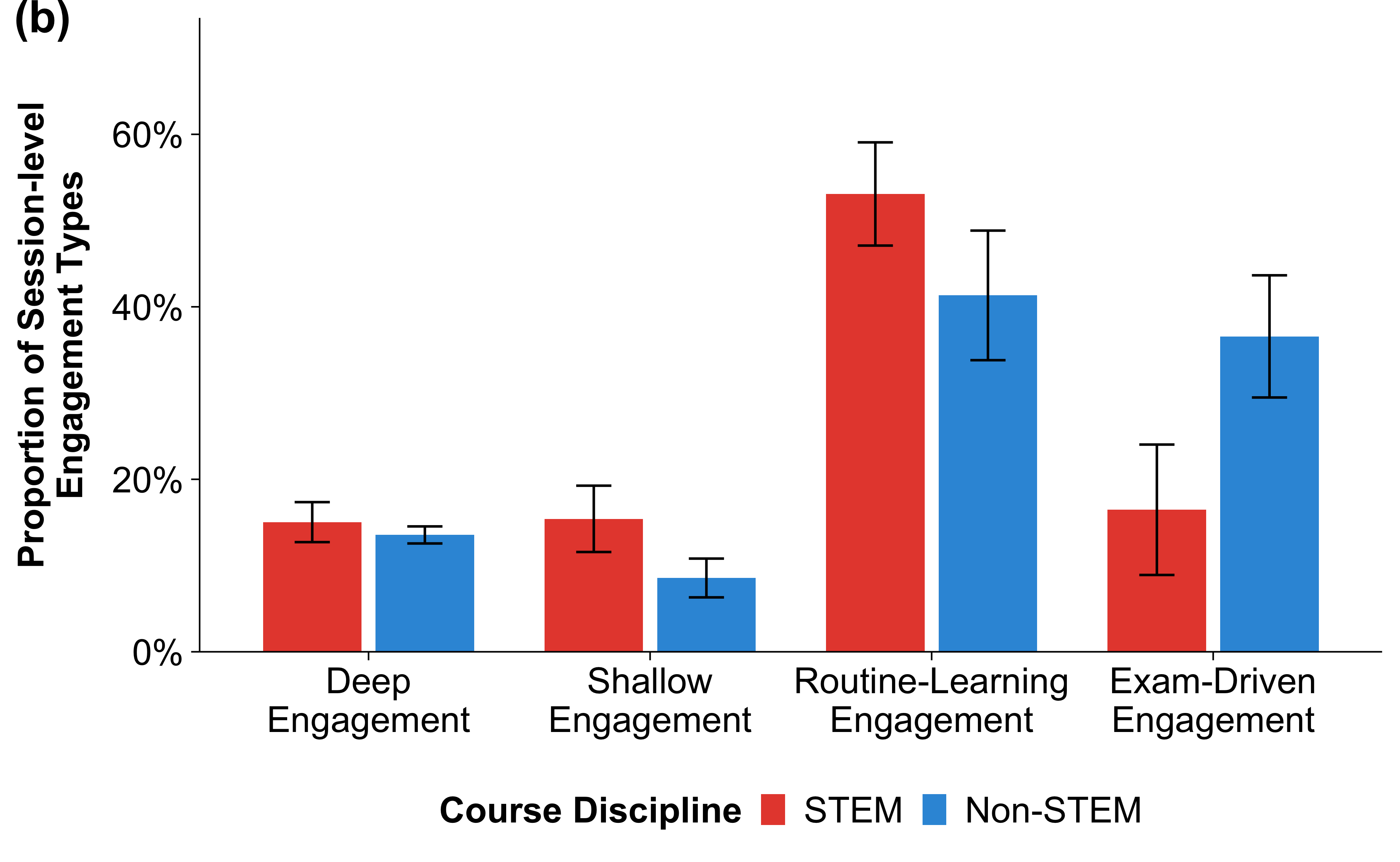}
    \caption{Distribution of Session-level Engagement Types by Institution Selectivity (a) and Course Discipline (b). Error bars show 95\% CIs with cluster-robust SEs.}
    \label{fig:engagement_context}
\end{figure*}

To explore how session-level engagement types varied across instructional contexts, we compared the proportion of each engagement type across institution selectivity and course discipline (Figure~\ref{fig:engagement_context}). 
In highly selective universities, the proportion of deep engagement and routine-learning engagement was significantly higher (deep engagement: 19.36\% vs.\ 12.42\%, $t(199) = 5.75$, $p < 0.01$; routine-learning engagement: 58.88\% vs.\ 40.41\%, $t(199) = 4.42$, $p < 0.01$), while the proportion of exam-driven engagement was lower (10.65\% vs.\ 36.98\%, $t(199) = -5.91$, $p < 0.01$). This suggests that students at highly selective universities may have engaged more deeply with the GenAI Tutor and relied on it more for routine learning rather than exam preparation.
Second, STEM courses showed a significantly higher proportion of shallow engagement (15.41\% vs.\ 8.56\%, $t(199) = 3.03$, $p < 0.01$) and routine-learning engagement (53.08\% vs.\ 41.32\%, $t(199) = 2.41$, $p = 0.02$), alongside a lower proportion of exam-driven engagement (16.47\% vs. 36.57\%, $t(199) = -3.81$, $p < 0.01$) compared to non-STEM courses. This pattern may suggest that STEM courses leveraged the GenAI Tutor for ongoing coursework support, whereas students in non-STEM courses were more likely to turn to the GenAI Tutor during exam preparation.

\subsection{Engagement Patterns at the Student Level}
\label{sec: Engagement Patterns}

\begin{figure}[htbp]
    \centering
    \includegraphics[width=\linewidth]{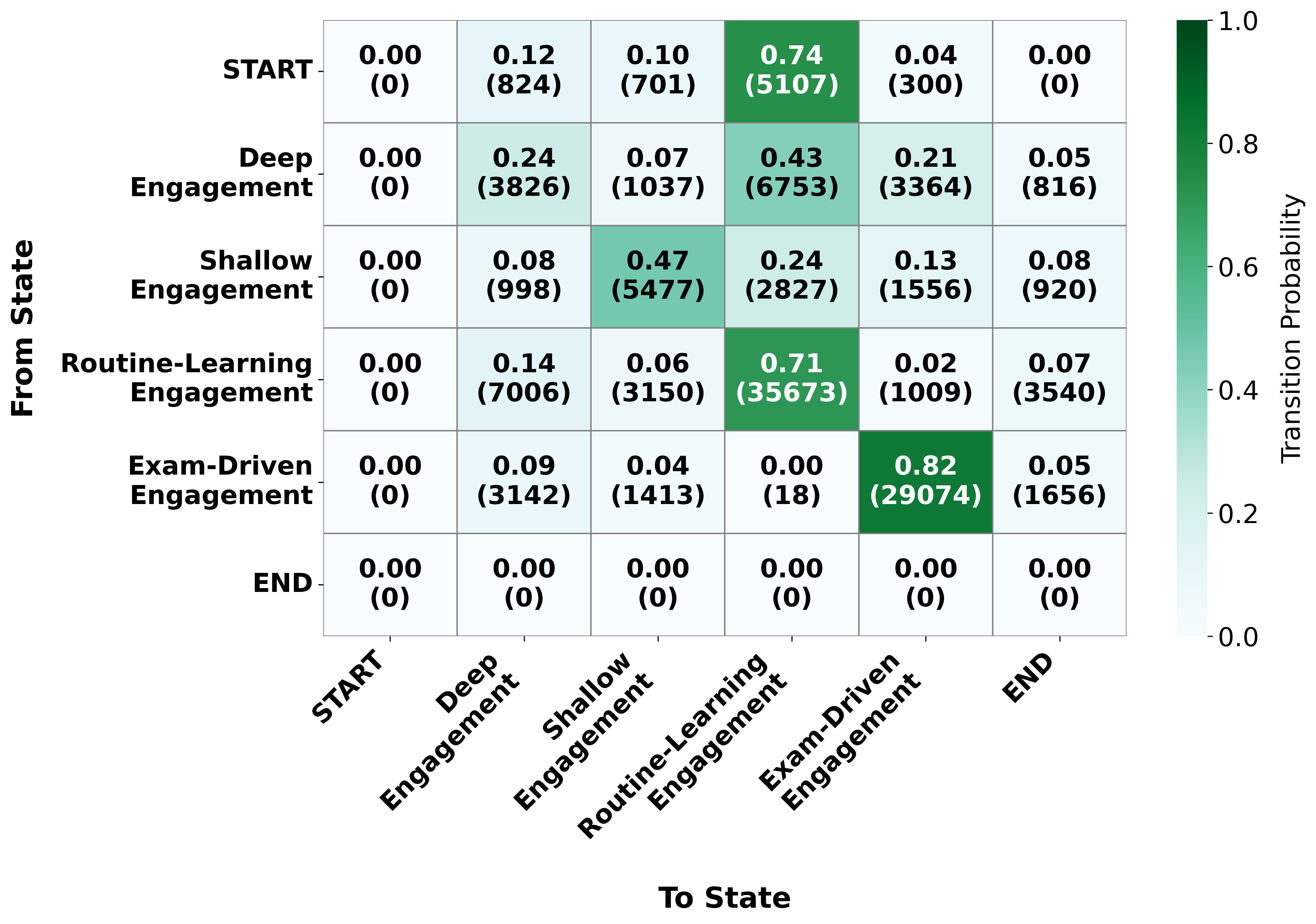}
    \caption{Transition Probability Matrix of Student Engagement Types. Each cell reports the conditional probability of transitioning from state $i$ (row) to state $j$ (column), with the absolute number of sequences shown in parentheses.}
    \label{fig:transition_prob}
\end{figure}

To answer the third research question, we built on the session-level clustering results and examined how students transition across engagement types as they interact with the GenAI Tutor (Figure~\ref{fig:transition_prob}). Four engagement patterns emerged:

1. \textbf{Deep Engagement Cycle}: This engagement pattern was characterized by repeated deep engagement (TP = 0.24), along with flexible transitions to routine-learning engagement (TP = 0.43) and exam-driven engagement (TP = 0.21). In contrast, deep engagement was unlikely to transition to shallow engagement (TP = 0.07). This indicates that students with deep engagement habits were likely to use the GenAI Tutor for both routine learning and exam preparation purposes throughout the semester and were unlikely to develop shallow engagement habits. 

2. \textbf{Shallow Engagement Cycle}: This engagement pattern was characterized by repeated shallow engagement (TP = 0.47), with very limited transitions to other engagement types (deep engagement: TP = 0.08; routine-learning engagement: TP = 0.24; exam-driven engagement: TP = 0.13). This suggests that students with shallow engagement habits were likely to stay in the surface use of the GenAI Tutor. 

3. \textbf{Routine-Learning Engagement Cycle}: This engagement pattern was the most common, as characterized by repeated routine-learning sessions (Transition Probability = 0.71). These sessions had a high probability of transitioning into deep engagement (TP = 0.14) compared to transitioning into other engagement types. This suggests that routine learning with the GenAI Tutor may have been sustained and intertwined with deep engagement over time. 
    
4. \textbf{Exam-Driven Engagement Cycle}: This engagement pattern was characterized by a high repetition of exam-driven engagement (TP = 0.82) and occasional transitions to deep engagement (TP = 0.09). This suggests that exam-focused use of the GenAI Tutor was intensive but still involved deep engagement, likely because students needed to genuinely learn to prepare for their exams.

\begin{figure*}[t]  
    \centering
    \includegraphics[width=0.48\textwidth]{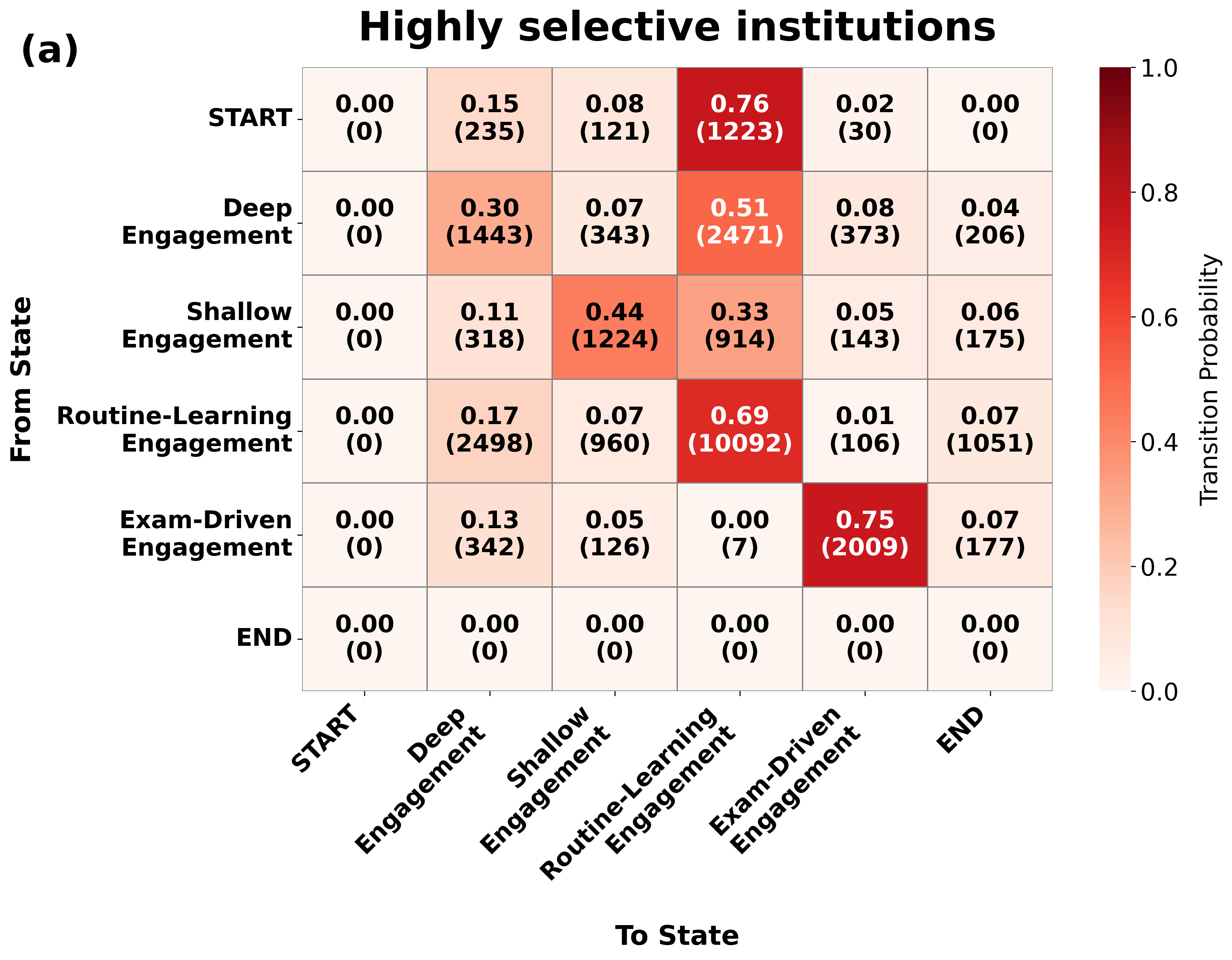}
    \hfill
    \includegraphics[width=0.48\textwidth]{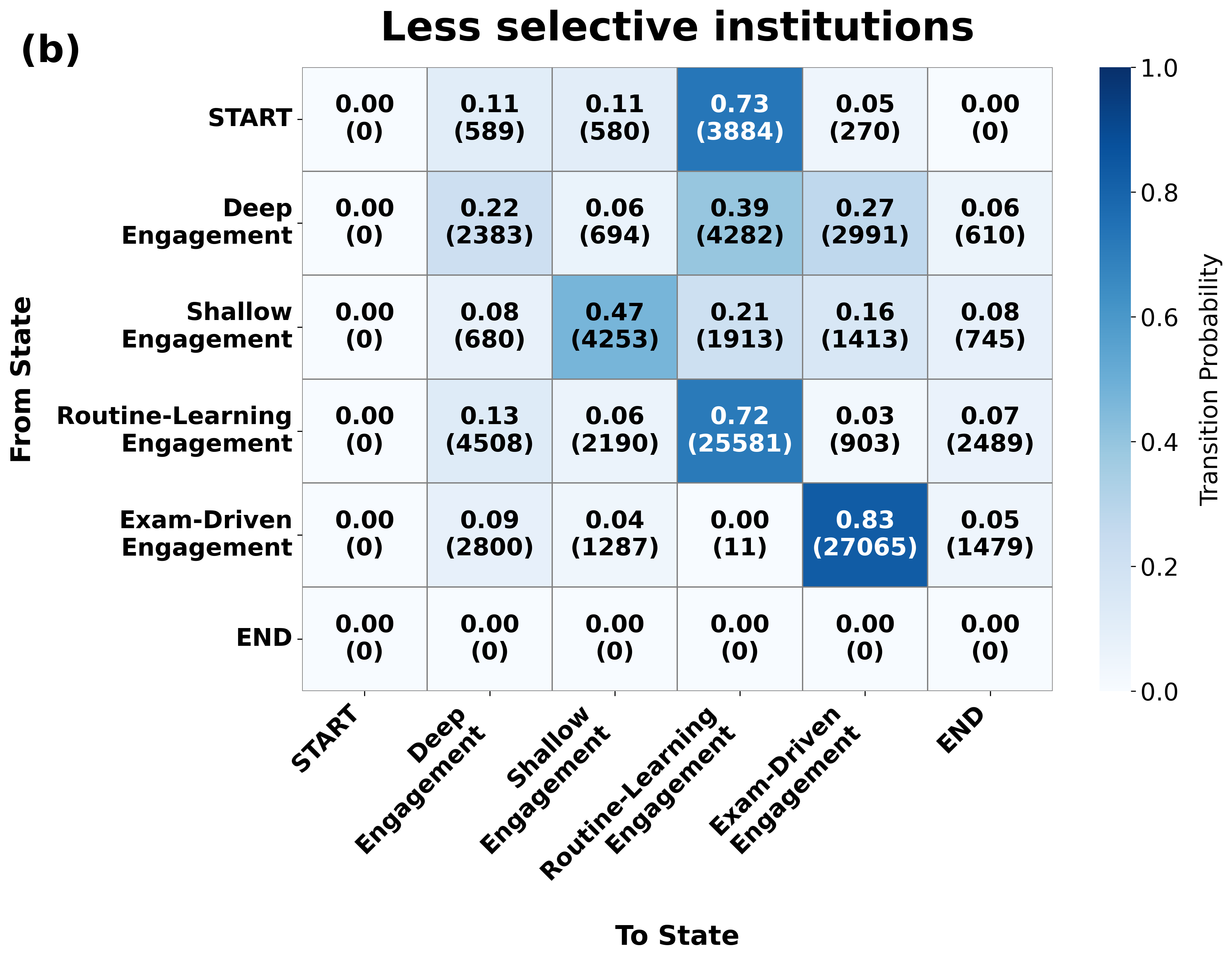}
    \caption{Transition Probability Matrices by Institution Selectivity: Highly Selective (a) vs. Less Selective (b)}
    \label{fig:transitions_institution}
\end{figure*}

Then we explored contextual variations in the engagement patterns of students by comparing transition probability matrices between subgroups. 
When students started deep engagement sessions, those from highly selective institutions were more likely to loop within deep engagement (TP = 0.30) compared to those from less selective institutions (0.22 for less selective institutions) (Figure~\ref{fig:transitions_institution}). They are also more likely to transition to routine-learning engagement (TP = 0.51 for highly selective institutions vs. TP = 0.39 for less selective institutions), rather than exam-driven engagement (TP = 0.08 vs. 0.27). When students started shallow engagement sessions, those from highly selective institutions were more likely to transition to routine-learning engagement (TP = 0.33 vs. 0.21) rather than exam-driven engagement (TP = 0.05 vs. 0.16), and less likely to loop within shallow engagement (TP = 0.44 vs. 0.47). These patterns suggest that students from highly selective institutions tended to integrate both deep and shallow engagement with routine learning, whereas students from less selective institutions were more likely to use the GenAI Tutor for shallow use, and then transition into exam preparation. Additionally, we found that once exam-driven engagement was initiated, students in less selective institutions were more likely to stay in the cycle (TP = 0.83 for less selective institutions vs. 0.75 for highly selective institutions), further suggesting that they have a stronger need to use the GenAI Tutor for exam preparation than students from highly selective institutions. 

\begin{figure*}[t]
    \centering
    \includegraphics[width=0.48\textwidth]{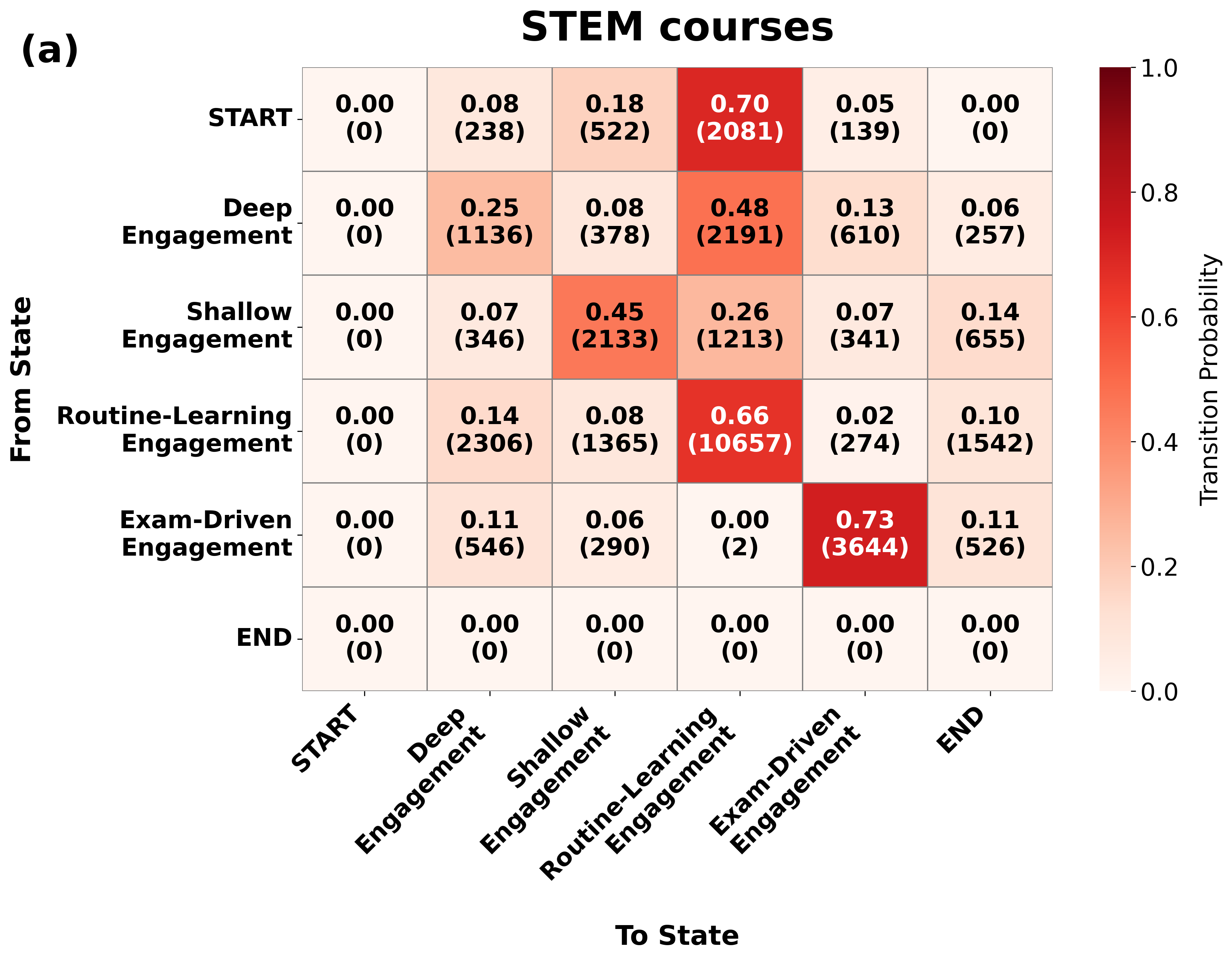}
    \hfill
    \includegraphics[width=0.48\textwidth]{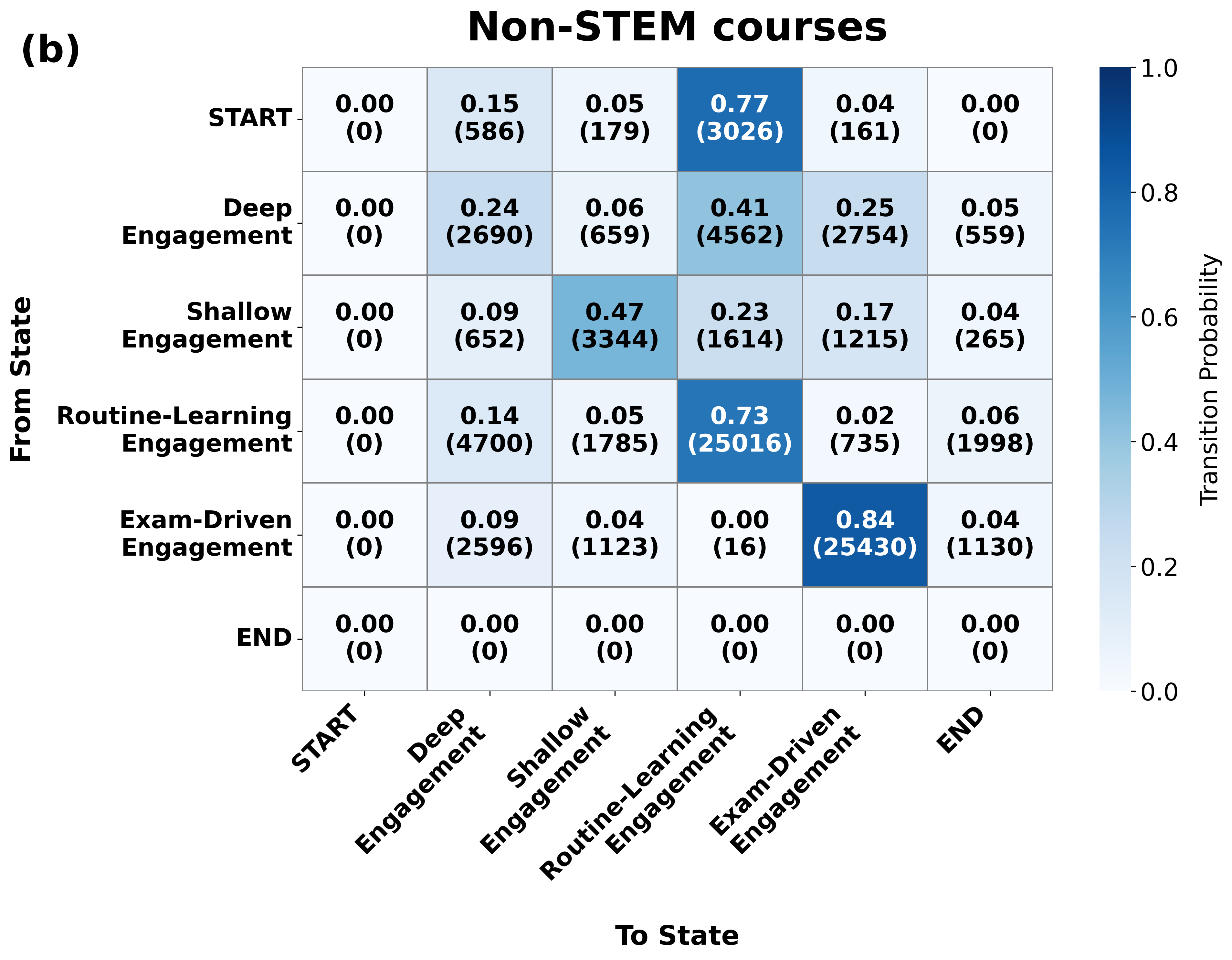}
    \caption{Transition Probability Matrices by Course Discipline: STEM (a) vs. Non-STEM (b)}
    \label{fig:transitions_discipline}
\end{figure*}

When it comes to disciplinary differences (Figure~\ref{fig:transitions_discipline}), for STEM students, they were more likely to initiate shallow engagement (TP = 0.18 for STEM courses vs. 0.05 for non-STEM courses), and among those who did, they tend to end the session (TP = 0.14 vs. 0.04). In comparison, for non-STEM students, although they were less likely to initiate shallow engagement, once they did so, they were more likely to transition to exam-driven engagement (TP = 0.17 for non-STEM courses vs. 0.07 for STEM courses). These patterns showed that shallow engagement appeared at different stages of learning between STEM and non-STEM. For STEM students, it was more closely tied to routine learning, but for non-STEM students, it appeared more often during exam preparation. Additionally, when students started deep engagement sessions, these sessions more often transitioned to routine-learning engagement for STEM students (TP = 0.48 for STEM courses vs. 0.41 for non-STEM courses) and transitioned to exam-driven engagement for non-STEM students (TP = 0.25 for non-STEM courses vs. 0.13 for STEM courses). Therefore, deep engagement was more likely to occur as part of routine learning for STEM students and as part of exam preparation for non-STEM students.

\section{Discussion}

This study examined student engagement with a GenAI Tutor on a commercial learning management system, using interaction data from more than 10,000 students across 200 classes in ten post-secondary institutions. We designed a two-stage analytical pipeline that first characterized engagement types at the conversation session level and then examined engagement patterns at the student level, drawing on behavioral, cognitive, and temporal characteristics of engagement. Overall, GenAI Tutor adoption was widespread, with 60.8\% students using the tool at least once during the semester, and then a decreasing subset of students continued to use the tool more intensively as the semester progressed. At the session level, we identified four distinct engagement types: Deep Engagement, Shallow Engagement, Routine-Learning Engagement, and Exam-Driven Engagement. At the student level, students typically iterated within engagement types but also transitioned across them over time. In particular, both session-level engagement types and student-level engagement patterns varied systematically across institutional selectivity and course discipline. Together, these findings reveal substantial nuances in how students adopt, use, and interact with GenAI Tutors in practice.
 
A central debate around the use of GenAI in education is that students may use these tools to bypass learning, such as asking for direct answers \cite{giannakos2025promise, kasneci2023chatgpt, yan2024promises}. Although such concerns have been documented in survey studies \cite{baek2024chatgpt, reiter2025student} and small-scale experiments \cite{fan2025beware, stadler2024cognitive}, little is known about how prevalent these behaviors are or how they evolve over time in authentic classrooms. Our findings provide novel insights into this issue. We show that shallow engagement, as commonly characterized by copy-pasting behaviors and direct answer requests, did occur but was not the dominant mode of GenAI Tutor use. Furthermore, shallow engagement did not dominate in high-stakes situations: During exam preparation, engagement with the GenAI Tutor exhibited the lowest levels of copy-pasting. This suggests that students often turn to GenAI Tutor for broad learning support rather than simple answer extraction. 

It should be noted that our identification of shallow engagement is based on interaction indicators, and further validation based on outcome-based measures and other process-based measures is needed \cite{saint2018detecting}. Moreover, our analyses focused on classes and institutions with a high adoption of GenAI Tutor, and these estimates may not fully reflect settings with a low use of GenAI Tutor. Additionally, students who choose to use the GenAI Tutor may have a higher motivation to learn compared to students who seek help from general GenAI tools. We do not have any data that could indicate students' usage and interaction with general GenAI tools. Overall, our findings suggest that concerns about learning avoidance are warranted but incomplete, and that student engagement with GenAI Tutors reflects a more nuanced mix of learning-supportive and learning-avoiding behaviors than commonly assumed.

Furthermore, our contextual analyses echo concerns about educational inequities related to GenAI use \cite{holmes2023guidance}. Prior research on scaling new educational technologies has often documented digital divides in access and usage intensity across learning contexts \cite{hannan2025widening, holt20245,kizilcec2017closing}. However, we did not observe such a divide in \textit{whether} or \textit{how much} GenAI Tutor was used by institutional selectivity, but in \textit{how students engaged with} the GenAI Tutor. At the session level, deep engagement and routine-learning engagement were more prevalent in highly selective institutions, suggesting that GenAI Tutor was used more for day-to-day learning rather than short-term assignment completion or exam cramming in these universities. At the student level, engagement differences among students accumulated over time: students with shallow engagement tended to remain confined to this mode over time, with a lower likelihood of transitioning into other engagement types, but students who engage deeply moved more flexibly across engagement types. This phenomenon of engagement disparities became even more pronounced when institutional selectivity was taken into account: Over time, students from highly selective institutions iterated within deep engagement more, and transitioned across other engagement types more adaptively. One possible explanation of such disparity is differences in students’ self-regulated learning skills: Those with stronger self-regulation may adapt how they engage with learning resources, but those with weaker self-regulation are more likely to persist in less effective learning strategies   \cite{cristea2025dynamics, saqr2023intense}. Together, these findings suggest a potential shift in equity challenges surrounding GenAI Tutoring, where inequities may be obscured by whom or how much they use these tools, but revealed in how they use these tools. 

Another key finding points to disciplinary limits in scaling GenAI Tutors for productive engagement. GenAI tools are commonly described as domain-agnostic, particularly compared to ITSs built around domain-specific learning modules \cite{mousavinasab2021intelligent}. However, our results showed that when GenAI Tutors were used in authentic classrooms, student engagement varied substantially between STEM and non-STEM disciplines. In STEM courses, shallow engagement with the GenAI Tutor often occurred together with routine learning and ended after a single session, whereas in non-STEM courses, shallow engagement occurred more often during exam preparation and was more likely to persist across multiple sessions. These engagement differences likely reflect disciplinary variations in instructional design and assessment structures. STEM courses often design weekly closed-ended problems. Therefore, students may have turned to GenAI for brief, task-oriented use. In contrast, non-STEM courses more commonly involve open-ended assignments for weekly instructions and take-home projects during exam periods. Therefore, students' reliance on the GenAI Tutor may have happened more iteratively and later in the semester. In summary, our findings suggest that scaling GenAI Tutors for productive engagement depends not only on advances in backend model generalizability but also on careful, discipline-sensitive design of how these tools are embedded and used in classrooms. This insight aligns with prior work showing that disciplinary differences pose an important barrier when digital technologies are introduced into university classrooms \cite{mercader2020university, qu2024disciplinary}. GenAI Tutors should not be viewed as a one-size-fits-all solution, but require discipline-sensitive design to support effective use at scale.

Our study points to several directions for future research. First, future work should examine the instructional and motivational factors underlying the engagement patterns we have observed, including how instructional design, assessment structures, and students’ learning goals shape engagement with GenAI Tutors. Such insights can inform instructional interventions and GenAI Tutor designs that promote more productive engagement and help mitigate emerging inequities.  Second, it will be informative to combine engagement patterns and other GenAI interaction indicators with performance outcomes to evaluate the downstream consequences of different engagement behaviors. This will help identify which engagement patterns are most supportive of learning and address the growing need to disentangle observed performance from genuine learning in the presence of GenAI assistance \cite{yan2025distinguishing}. Finally, we would be interested in applying the analytical approach we have developed in other settings, and inform more responsible and context-sensitive scaling of GenAI products.

\section{Conclusion}

This study provides a large-scale, multi-institutional view of how students engage with a GenAI Tutor in authentic post-secondary classrooms. By analyzing interaction logs at both the conversation session and student levels, we show that student engagement with GenAI Tutors is heterogeneous, dynamic over time, and shaped by disciplinary and institutional contexts. Although adoption is widespread, meaningful differences emerge in the way students use these tools, including shallow, deep, routine-learning, and exam-driven engagement patterns. These findings highlight the importance of moving beyond access and usage intensity to examine engagement quality when evaluating GenAI Tutors.

\section{Acknowledgments}

We would like to thank Wei Wang and Yun Wang from XuetangX and Shuangshuang Guo from the Online Education Center, Tsinghua. This study is funded by the MOE Research Center for Online Education. Liu gratefully acknowledges financial support from the National Natural Science Foundation of China (72342032,72222005). Ga\v{s}evi\'{c} would like to acknowledge the support from the Australian Government through the Australian Research Council (DP240100069 and DP220101209) and the Jacobs Foundation (CELLA 2 CERES). The views expressed here are those of the authors and do not necessarily reflect the views of the company or the funders.

\printbibliography


\end{document}